\documentclass[aps,prl,nofootinbib,showpacs,preprint,floatfix]{revtex4}
\usepackage{latexsym,epsfig,graphicx,axodraw}

\newcommand{\vs}[1]{\rule[- #1 mm]{0mm}{#1 mm}}
\newcommand{\qqbar}{\bar q^{\!\!\!\!\!\textsuperscript{\tiny{(~~)}}}}
\newcommand{\ppbar}{\bar p^{\!\!\!\!\!\textsuperscript{\tiny{(~~)}}}}
\newcommand{\bbbar}{\bar b^{\!\!\!\!\!\textsuperscript{\tiny{(~~)}}}}
\newcommand{\bfig}{\begin{center}\begin{picture}}
\newcommand{\efig}[1]{\end{picture}\\{\small #1}\end{center}}

\newcommand{\wlin}[2]{\DashLine(#1)(#2){2.5}}

\newcommand{\glin}[3]{\Gluon(#1)(#2){2}{#3}}

\newcommand{\sof}{\SetOffset}
\newskip\humongous \humongous=0pt plus 1000pt minus 1000pt

\newif\ifdtup

\newcommand{\eq}{\vs{2}\begin{equation}}
\newcommand{\be}{\vs{2}\begin{equation}}
\newcommand{\en}{\\[2mm]\end{equation}}
\newcommand{\bea}{\begin{eqnarray}}
\newcommand{\ena}{\end{eqnarray}}
     
\newcommand{\lapprox}{%
\mathrel{%
\setbox0=\hbox{$<$}
\raise0.6ex\copy0\kern-\wd0
\lower0.65ex\hbox{$\sim$}
}}
\newcommand{\gapprox}{%
\mathrel{%
\setbox0=\hbox{$>$}
\raise0.6ex\copy0\kern-\wd0
\lower0.65ex\hbox{$\sim$}
}}

\begin{document}
\bibliographystyle{plain}
\begin{titlepage}
\begin{flushright}
{
UG-FT-185/05 \\
CAFPE-55/05 \\
July 2005\\}
\end{flushright}

\begin{center}
{\bf{\Large{\bf{NLO forward-backward charge asymmetries in 
$p \bar p^{\!\!\!\!\!\textsuperscript{\tiny{(\,~~)}}} \rightarrow l^-l^+j$ 
production at large hadron colliders}}}}
\\[0.5cm]
\indent
{\bf F. del Aguila$^a$, Ll. Ametller$^b$ and R. Pittau$^a$
\footnote{On leave of absence from Dipartimento di Fisica Teorica, 
Torino and INFN Sezione di Torino, Italy.}}
\\[0.5cm]
{$^a$ {\it Departamento de F{\'\i}sica Te\'orica y del Cosmos 
and Centro Andaluz de F{\'\i}sica de Part{\'\i}culas Elementales (CAFPE),
Universidad de Granada, 
E-18071 Granada, Spain.}\\[.1cm]
$^b$ {\it Departament de F{\'\i}sica i Enginyeria Nuclear,
Universitat Polit\`ecnica de Catalunya,\\
E-08034 Barcelona, Spain.}
}
\\[0.5cm]
\vfill
{\bf Abstract} \\
\end{center}
{We consider the next-to-leading order corrections, $\cal O(\alpha _{\rm s})$, 
to forward-backward charge asymmetries for lepton-pair production in 
association with a large transverse momentum jet at large hadron colliders. 
We find that 
the leading order results are essentially confirmed. Although experimentally 
challenging and in practice with large backgrounds, these observables 
could provide a new determination of the weak mixing angle   
$\sin ^2 \theta _{\rm eff}^{\rm lept}(M_Z^2)$ with a statistical 
precision for each lepton flavour of $\sim  10^{-3}\ (7\times 10^{-3})$ at LHC (Tevatron), and 
if $b$ jets are identified, of 
the {\it b} quark $Z$ asymmetry ${\rm A}_{\rm FB}^b$ with a 
statistical precision of $\sim  2\times 10^{-3}\ (4\times 10^{-2})$ 
at LHC (Tevatron).
}
\noindent
\vfill
\begin{center}
 \\[0.2mm]
{\bf Keywords:} 
\begin{minipage}[t]{12cm}
Hadron-induced high- and super-high-energy interactions,
Gauge bosons. 
\end{minipage}
\end{center}
\end{titlepage}

\renewcommand{\thefigure}{\arabic{figure}}

\section{INTRODUCTION}
The large cross sections for gauge boson production at the 
Fermilab Tevatron and the CERN Large Hadron Collider (LHC) 
might give a chance to determine the electroweak parameters 
with high precision \cite{Tevatron,Haywood00}.
In practice, the experimental challenge is very demanding, 
but at any rate pursuing such measurements will help to 
disentangle the strong physics contributing to the 
different processes. 
In this paper we calculate the forward-backward charge 
asymmetries of lepton pairs in events with a 
large transverse momentum jet 
$p \ppbar \rightarrow Z, \gamma ^* + j \rightarrow e^-e^+ + j$
\footnote{Throughout the paper we will explicitly refer to the 
$e^-e^+ (+ j)$ decay channel. The same analysis applies to 
$l^-l^+ + j$ production with $l = \mu$.}
at next-to-leading order (NLO), 
$\cal O(\alpha _{\rm s})$ corrections. 
We make use of the Monte Carlo program 
MCFM v4.1 \cite{MCFM}, which includes the necessary processes 
at this order, and of ALPGEN \cite{ALPGEN}.
The particularly interesting case of a final $b$ jet is 
discussed in detail. 
We find that the leading order (LO) predictions 
\cite{Aguila02} are essentially confirmed.  

Electron-positron pair production 
$p  \ppbar \rightarrow Z, \gamma ^* \rightarrow e^-e^+$ 
has a large cross section at large hadron colliders, and as it 
is sensitive to the presence of vector and axial-vector fermion 
couplings to neutral gauge bosons, in principle allows for their 
precise measurement. 
A prime example is the determination of the effective weak mixing 
angle $\sin ^2\theta _{\rm eff}^{\rm lept}$, that enters in their 
definition,  
the optimum observable being the forward-backward charge asymmetry 
of the lepton pairs ${\rm A}_{\rm FB}$ \cite{Rosner89,Fischer95}.
Indeed, the tree level Drell-Yan 
\footnote{A list of Drell-Yan cross section measurements 
at Tevatron run I and II is given in Ref. \cite{Affolder01}.} 
parton process $q\bar q \rightarrow Z, \gamma ^* \rightarrow e^-e^+$ 
gives an asymmetric polar angle electron distribution relative 
to the initial quark, which also depends on the lepton pair 
invariant mass $M_{e^-e^+}$

At the Fermilab Tevatron Run I the Collider Detector at Fermilab (CDF) 
reported
an asymmetry ${\rm A}_{\rm FB}$ at the $Z$ peak of
$0.07 \pm 0.02$ \cite{Affolder01a}, in agreement 
with the Standard Model (SM) prediction. 
A new measurement of neutral gauge boson production in 
$p\bar p$ collisions at the upgraded Run II Fermilab 
Tevatron operated at $\sqrt s = 1.96$ TeV has been recently presented
by CDF, 
giving, at the $Z$ peak,  ${\rm A}_{\rm FB} = 0.07 \pm 0.03$, 
the statistical error 
being large because only an integrated luminosity of 72 pb$^{-1}$ 
has been analysed \cite{CDF05a}, less than a tenth of the 
luminosity collected so far.
\footnote{In Run I the asymmetry found for the last bin, 
$M_{e^-e^+} \in [300,600]$ GeV, deviated from the SM prediction 
\cite{Affolder01a}, what is not confirmed by Run II \cite{CDF05a}.}
A fit to these data where the quark and electron couplings to the 
$Z$ boson are expressed as a function of 
$\sin ^2\theta _{\rm eff}^{\rm lept}$ 
gives 
$\sin ^2\theta _{\rm eff}^{\rm lept} = 0.2238 \pm 0.0040 \pm 0.0030$, 
where the errors stand for statistics and systematics, respectively. 
This is far away from the estimate of the expected statistical precision 
$\sim 0.0005$ to be reached at Run II with an integrated luminosity of 
10 fb$^{-1}$ \cite{Baur98}. Even if the experiment 
is well understood \cite{CDF05a} and the 
theoretical calculations not ambiguous, 
it seems very hard to get rid of systematic uncertainties to the required 
level. In the following with a more exclusive process we will need a more 
demanding experimental performance. But we will stay on the very optimistic 
side, emphasizing what we may learn if we were only limited by statistics.
At LHC with an integrated luminosity of 
100 fb$^{-1}$ the weak mixing angle precision would be hopefully 
further improved by a factor 
$\sim 3$. This would be comparable 
to the current global fit precision, $0.00016$,
but for instance a factor 
$\sim 2$ better than the effective weak mixing angle precision obtained 
from the bottom forward-backward asymmetry at LEP and SLD, $0.00029$
\cite{LEP02}.

The associated production of a neutral gauge boson $V=Z, \gamma ^*$ 
and a jet $j$ 
has also a large cross section, especially at LHC, and can also allow 
for a precise determination of $\sin ^2\theta _{\rm eff}^{\rm lept}$. 
This NLO correction to $V$ production 
is a genuine new process when we require the 
detection of the extra jet. 
In particular, gluons can be also initial states, 
and the large gluon content of the proton at high energy 
tends to increase the $Vj$ production cross sections, 
although they stay almost one order of magnitude smaller than 
the corresponding $V$ cross sections. 
In Table \ref{table:2} of the section that collects our numerical
results we gather the  different LO and NLO contributions to 
$V(\rightarrow e^-e^+)j$ production 
at Tevatron, to be compared with the inclusive LO 
and NLO $V \rightarrow e^-e^+$ cross section for the same cuts, 
127 and 158 pb, respectively. 
\noindent At LHC we find for $e^-e^+$ production 685 and 745 pb, respectively, 
to be compared to the $e^-e^+ j$ cross section, 53 and 57 pb, in
Table \ref{table:2} below.
All the calculations throughout the paper have been performed 
with MCFM v4.1, and with ALPGEN when necessary.
They provide a good description of these processes at hadron 
colliders. For instance, the prediction at Tevatron for the ratio of the inclusive 
cross section for $p\bar p \rightarrow V~\bbbar$ to  
$p\bar p \rightarrow Vj$ production is,
 according to the results in Table \ref{table:2} below, 
0.020 to NLO 
\footnote{We neglect in this estimate the small fraction of events 
where $b$ and $\bar b$ combine into the same jet.}
(0.0096 to LO \cite{Aguila02}). 
\footnote{In apparent agreement with the NLO prediction $0.018 \pm 0.004$ 
by J.M. Campbell and Willenbrock quoted in Ref. \cite{D005}.} 
This has to be compared to the recent measurement 
of this ratio with the D0 detector $0.023 \pm 0.005$ \cite{D005},
obtained with a similar, but not identical, set of cuts. 

As pointed out in Ref. \cite{Aguila02} the forward-backward 
charge asymmetry of the lepton pairs can be measured 
in neutral gauge boson production with an accompanying jet 
either relative to a direction fixed by the initial state 
${\rm A}^{\rm CS}_{\rm FB}$ as in the inclusive neutral gauge boson 
production (Drell-Yan case), or relative to the final jet direction 
${\rm A}_{\rm FB}^j$. 
The former is adapted to obtain the asymmetry 
from $q \bar q$ events, and the latter 
from $g$ $\qqbar$ ones.
Both asymmetries give similar precision for 
$\sin ^2\theta _{\rm eff}^{\rm lept}$ at LHC but not at 
Tevatron, where the precision for ${\rm A}^{\rm CS}_{\rm FB}$ is almost 
one order of magnitude higher. However, in principle 
${\rm A}_{\rm FB}^j$ also 
allows for the measurement of flavour asymmetries. Thus, if 
we require the final jet to be a $b$ quark, we can make a 
new measurement of ${\rm A}_{\rm FB}^b$. 
This is especially interesting given 
its observed deviation at the $Z$ pole from the SM prediction, 
$3\ \sigma$ \cite{LEP02}. 
However, to approach a similar precision will be a very 
demanding experimental challenge because we have not only to 
identify the heavy quark but to measure its charge. 
Being very optimistic the corresponding effective weak mixing angle 
precision to be in principle expected at LHC, 
$\sim 10^{-3}$, is already lower than the one reported by LEP and SLD, 
$2.9\times 10^{-4}$, but similar to the difference between the central 
values resulting from ${\rm A}_{\rm FB}^b$ at the $Z$ pole
and the global fit to all data \cite{LEP02}.

In the following we study the LHC and the Tevatron potentials 
in turn. 
First, we review the LO contributions to 
$p \ppbar \rightarrow Vj$ production and introduce 
the different asymmetries.  
Afterward we discuss 
the NLO corrections, paying special attention to the case of 
a final $b$ jet.
Finally, we present the numerical results and draw our conclusions. 

\section{LO processes and forward-backward charge asymmetries}
\begin{figure}[ht]
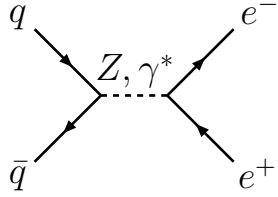

\bfig(100,70)
\SetWidth{1}
\SetScale{1}
\sof(0,0)
  \ArrowLine(0,50)(25,25)
  \ArrowLine(25,25)(0,0)
  \wlin{25,25}{50,25}
   \ArrowLine(50,25)(75,50)
   \ArrowLine(75,0)(50,25)
   \Text(38,28)[b]{{\large $Z,\gamma^\ast$}}
   \Text(-3,50)[rb]{{\large $q$}}
   \Text(-3,0)[rt]{{\large $\bar q$}}
   \Text(78,52)[lb]{{\large $e^-$}}
   \Text(77,1)[lt]{{\large $e^+$}}
\efig{}
\caption{\label{fig:01}
LO $q \bar q$ contribution to the Drell-Yan process 
in Eq. \ref{withoutj}.}
\end{figure}

\begin{figure}[th]
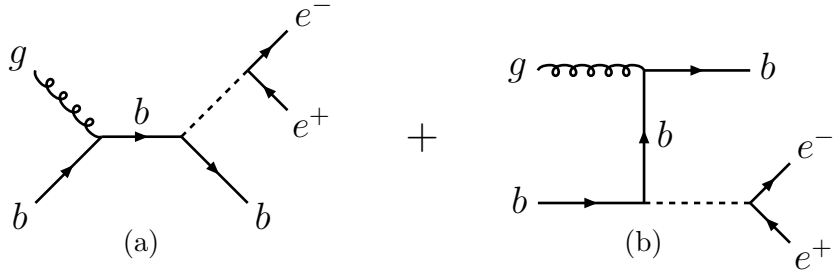

\bfig(100,105)
\SetWidth{1}
\SetScale{1}
\sof(-90,20)
\Text(40,-10)[t]{(a)}
  \glin{25,25}{0,50}{4}
  \ArrowLine(0,0)(25,25)
  \ArrowLine(25,25)(55,25)
  \wlin{80,50}{55,25}
  \ArrowLine(55,25)(80,0)
   \ArrowLine(80,50)(95,65)
   \ArrowLine(95,35)(80,50)
   \Text(-3,50)[rb]{{\large $g$}}
   \Text(-3,0)[rt]{{\large $b$}}
   \Text(40,30)[b]{{\large $b$}}
   \Text(83,0)[lt]{{\large $b$}}
   \Text(98,67)[lb]{{\large $e^-$}}
   \Text(97,36)[lt]{{\large $e^+$}}
   \Text(140,30)[lt]{{\Large $+$}}
\sof(100,20)
\Text(40,-10)[t]{(b)}
  \glin{0,50}{40,50}{5}
  \ArrowLine(40,50)(80,50)
  \ArrowLine(0,0)(40,0)
  \wlin{40,0}{80,0}
  \ArrowLine(40,0)(40,50)
   \ArrowLine(80,0)(95,15)
   \ArrowLine(95,-15)(80,0)
   \Text(-4,50)[r]{{\large $g$}}
   \Text(-4,1)[r]{{\large $b$}}
   \Text(45,26)[l]{{\large $b$}}
   \Text(84,52)[l]{{\large $b$}}
   \Text(98,17)[lb]{{\large $e^-$}}
   \Text(97,-14)[lt]{{\large $e^+$}}
\efig{}
\caption{\label{fig:02}
LO $gb$ contributions to the process in Eq. \ref{withj} for $j= b$.}
\end{figure}

Let us thus compare the processes 
\begin{eqnarray}
\label{withoutj}
p \ppbar \rightarrow Z, \gamma ^* \rightarrow e^-e^+ 
\end{eqnarray}
and
\begin{eqnarray}
\label{withj}
p \ppbar \rightarrow Z, \gamma ^* + j \rightarrow e^-e^+ + j
\end{eqnarray}
at LO and define the different forward-backward asymmetries 
we are interested in.
Fig. \ref{fig:01} shows the LO diagram contributing to the 
Drell-Yan process in Eq. \ref{withoutj}.
In the absence of gluonic radiation the transverse momentum of the
exchanged vector boson is zero. Therefore, one must, for instance, 
expect that the direction of the initial quark state and the final 
$e^-$ are correlated.
When the initial quark line emits gluons, such correlations, 
although still present, tend to diminish because of the transverse 
momentum $p_t$ acquired by the $e^- e^+$ system.
If one of those additional gluons is hard enough and is emitted in 
the central region of the detector, it gives rise to an extra jet 
resulting into the process in Eq. \ref{withj}. 
However, allowing for an extra jet obliges to consider new 
subprocess initiated by a gluon and a(n) (anti)quark.
We show in Fig. \ref{fig:02} the relevant tree level diagrams 
in the simple case $j= b$.
In diagram (b) the decay products of the $Z,\gamma^\ast$
system know very little about the direction of the final state $b$, 
because of the initial state gluon
that also connects to the $b$ fermionic line. Instead, in diagram (a), 
the $p_t$ of the $b$ quark exchanged in the $s$-channel
is zero, therefore one expects correlations between the 
final state leptons and the direction of the $b$-jet
\footnote{Analogous considerations also hold in the general
case with $j \ne b$.} \cite{Aguila02}.
An optimal observable to quantify such correlations 
for the process of Eq. \ref{withj} is a forward-backward asymmetry:
\begin{eqnarray}
{\rm A}_{\rm FB} = \frac {F-B}{F+B}\,, 
\end{eqnarray}
with
\begin{eqnarray}
F = \int _0^1 \frac{{\rm d}\sigma}
{{\rm d}\cos \theta }
{\rm d}\cos \theta , \,\,\,  
B = \int _{-1}^0 \frac{{\rm d}\sigma}
{{\rm d}\cos \theta }
{\rm d}\cos \theta \,. \nonumber
\end{eqnarray}
One can consider two possible angles:
\begin{eqnarray}
\cos \theta _{\rm CS} &=&
\frac{2 (p_z^{e^-}E^{e^+}-p_z^{e^+}E^{e^-})}
{\sqrt {(p^{e^-} + p^{e^+})^2}
\sqrt {(p^{e^-} + p^{e^+})^2 + (p^{e^-}_T+p^{e^+}_T)^2 }}\,, \nonumber \\
\cos \theta _j &=& 
\frac{(p^{e^-}-p^{e^+})\cdot p^j}
{(p^{e^-}+p^{e^+})\cdot p^j}\,, \nonumber 
\label{costheta} 
\end{eqnarray}
where the four-momenta are measured in the 
laboratory frame and $p^{\mu}_T \equiv (0,p_x,p_y,0)$.
The Collins-Soper angle \cite{Collins77} $\theta _{\rm CS}$ is, 
on average, the angle between $e^-$ and the initial quark direction,
while $\theta _j$ is the angle between 
{$e^-$} and the direction opposite to the jet in the 
{$e^-e^+$} rest frame \cite{Aguila02}.
From the previous discussion it should be clear that the former choice 
is adapted to the 
$q\bar q$ collisions and the latter to the $g$ $\qqbar$ ones.

Different asymmetries can be defined, according to 
the scheme given in Table \ref{table:1a}.
\begin{table}[ht]
\begin{center}
\begin{tabular}{|c||c|l|}
\hline
Collider & Asymmetry & ~~~~~~~~~~~Definition \\ \hline \hline
$p \bar p $ & ${ A^{\rm CS}_{\rm FB}}$ & 
$\cos \theta = \cos \theta_{\rm CS}$    \\
$p      p $ & ${ A^{\rm CS}_{\rm FB}}$& $\cos \theta = \cos \theta_{\rm CS} 
 \times \frac{|p_z^{e^+}+p_z^{e^-}+p_z^j|}{p_z^{e^+}+p_z^{e^-}+p_z^j}$ 
       \\ \hline
$p \bar p $ & ${ A^{j}_{\rm FB}}$ & $\cos \theta = \cos \theta _j 
     \times \frac{|p_z^{e^+}+p_z^{e^-}+p_z^j|}{p_z^{e^+}+p_z^{e^-}+p_z^j}$ \\ 
$p      p $ & ${ A^{j}_{\rm FB}}$ & $\cos \theta = \cos \theta _j$ \\ \hline
$p \bar p $ & ${ A^{b}_{\rm FB}}$ & $\cos \theta = \cos \theta _j 
             \times (-{\rm sign}(Q_b))$ \\
$p      p $ & ${A^{b}_{\rm FB}}$ & $\cos \theta = 
\cos \theta _j\times (-{\rm sign}(Q_b))$ 
\\ \hline
\end{tabular}
\end{center}
\caption{\label{table:1a}
The definitions of the various asymmetries at the $pp$ and $p \bar p$
colliders.}
\end{table}
A comment is in order, with respect to the phases appearing in the Table.
In $pp$ colliders the quark direction is fixed by the 
rapidity of the jet plus the lepton pair. This implies defining 
$\cos \theta _{CS}$ with an extra sign factor 
$\frac{|p_z^{e^-}+p_z^{e^+}+p_z^j|}{p_z^{e^-}+p_z^{e^+}+p_z^j}$, 
as in the second line of Table \ref{table:1a} . 
On the other hand, in $p\bar p$ colliders there are produced as 
many quarks as antiquarks and $A^{j}_{\rm FB}$ vanishes unless 
some difference is made between them.  
Hence, $\cos \theta $ is defined with an extra sign factor 
$\frac{|p_z|}{p_z}$, $p = p^{e^-} + p^{e^+} + p^j$,
which corresponds to 
assume that the largest rapidity parton is a (anti)quark 
if it is along the (anti)proton direction. This explains the factor
in the third line of Table \ref{table:1a}.
Finally, because both in $pp$ and $p\bar p$ colliders are produced  
as many $b$ as $\bar b$, in order to obtain ${\rm A}^b_{\rm FB}$
one must use $\cos \theta_ j$  multiplied by 
a $+(-)$ sign for b (anti)quarks, $-{\rm sign}(Q_b)$ 
with $Q_b$ the $b$ charge. In practice this means detecting the
charge of the produced $b$ jet. 

Such asymmetries have been studied in detail, at LO, 
in Ref. \cite{Aguila02}.
What is important to realize is that, in order to get reliable predictions,
a priori small additional contributions 
must be carefully taken into account.
Let us study, in particular, the effect of
including radiative ${\cal O} (\alpha_s)$, NLO, corrections.
\section{NLO contributions}
In the following we discuss all contributions necessary 
to compute the process in Eq. \ref{withj} at the NLO.
The described structure is implemented in the MCFM code, that we 
used as it is in the case $j \ne b$.
 However, as already pointed out, the computation of $A^b_{\rm FB}$
requires disentangling $b$ from $\bar b$ final states. 
In MCFM this selection is not possible on an event by event basis, because  
$b$ and $\bar b$ contributions are summed up. Therefore, we modified 
the code to take this into account. In addition, we included part 
of the remaining real NLO contribution with the help of ALPGEN.
We find convenient to list the different contributions in the following, 
for the case $j= b$, because
this will allow us to discuss their relative size. 
Analogous considerations apply to the general case.

\begin{figure}[ht]
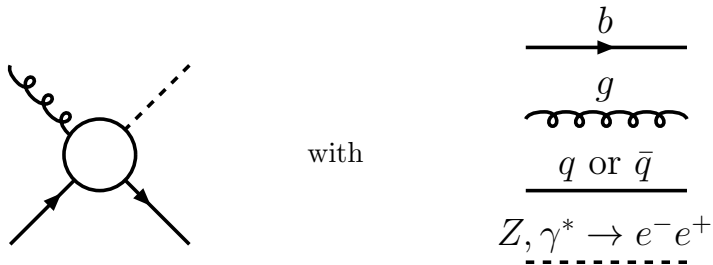

\bfig(100,90)
\SetWidth{1}
\SetScale{1.35}
\sof(-75,0)
  \glin{25,25}{0,50}{4}
  \ArrowLine(0,0)(25,25)
  \wlin{50,50}{25,25}
  \ArrowLine(25,25)(50,0)
  \GCirc(25,25){10}{1.0}
\SetScale{1}
  \Text(123,35)[]{\rm with}
\SetScale{1.35}
\sof(120,0)
  \ArrowLine(0,55)(45,55)   \Text(30,80)[b]{\large $b$}
       \glin{0,35}{45,35}{5}\Text(30,54)[b]{\large $g$}
       \Line(0,15)(45,15)   \Text(30,24)[b]{\large $q$ or $\bar q$}
       \wlin{0,-5}{45,-5}   \Text(30,-3)[b]{\large $Z,\gamma^\ast \to e^- e^+$}
\efig{}
\caption{\label{fig:nlovirt} 
NLO virtual $gb$ contributions.}
\end{figure}

The virtual contributions are drawn in Fig. \ref{fig:nlovirt},
together with the definition of all the graphical symbols and conventions  
used. In particular, we omit 
drawing explicitly the decay of the  $Z,\gamma^\ast$ system, but
we always understand $Z,\gamma^\ast \to e^- e^+$, and the blob
stands for the sum of all possible contributing Feynman diagrams.
For the case at hand, it means exchanging a virtual gluon
in all possible ways in the two diagrams of Fig. \ref{fig:02}.

\begin{figure}[ht]
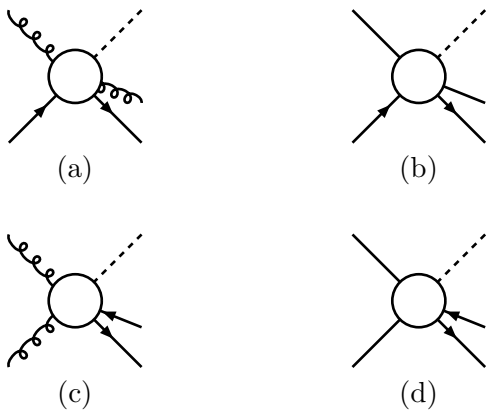

\bfig(100,155)
\SetWidth{1}
\SetScale{1}
\sof(-40,90)
\Text(25,-5)[t]{(a)}
  \glin{25,25}{0,50}{4}
  \ArrowLine(0,0)(25,25)
  \wlin{50,50}{25,25}
  \ArrowLine(25,25)(50,0)
  \glin{25,25}{50,15}{4} 
  \GCirc(25,25){10}{1.0}
\sof(90,90)
\Text(25,-5)[t]{(b)}
  \Line(25,25)(0,50)
  \ArrowLine(0,0)(25,25)
  \wlin{50,50}{25,25}
  \ArrowLine(25,25)(50,0)
  \Line(50,15)(25,25)
  \GCirc(25,25){10}{1.0}
\sof(-40,5)
\Text(25,-5)[t]{(c)}
  \glin{25,25}{0,50}{4}
  \glin{0,0}{25,25}{4}
  \wlin{50,50}{25,25}
  \ArrowLine(25,25)(50,0)
  \ArrowLine(50,15)(25,25)
  \GCirc(25,25){10}{1.0}
\sof(90,5)
\Text(25,-5)[t]{(d)}
  \Line(0,50)(25,25)
  \Line(0,0)(25,25)
  \wlin{50,50}{25,25}
  \ArrowLine(25,25)(50,0)
  \ArrowLine(50,15)(25,25)
  \GCirc(25,25){10}{1.0}
\efig{} 
\caption{\label{fig:nloreal}
NLO real $gb$ (a), 
 $\,\bar q^{\!\!\!\!\!\textsuperscript{\tiny{\rm (~~)}}}\,b$ 
 (b),
$gg$ (c), $q \bar q$ (d) contributions.}
\end{figure}

The real contributions are given in Fig. \ref{fig:nloreal}.
The shorter lines on the right part of the drawings means
that the corresponding outgoing partons are not seen because they
are too soft 
or too collinear to the ingoing or outgoing b quark, therefore
also leading to a final state 
formed by an $e^-e^+$ pair plus a jet containing a $b$ quark.
The NLO fully differential cross section ${\rm d} \sigma^{\rm NLO}$
is given by 
\begin{eqnarray}
\label{master}
{\rm d} \sigma^{\rm NLO} &=& \sum_{i,j} \int_0^1 dx_1 \int_0^1 dx_2~
      f^{\rm NLO}_i{(x_1)} f^{\rm NLO}_j{(x_2)}~ 
      {\rm d} \hat \sigma_{ij}^{\rm NLO}\,, \nonumber \\
i,j &=& g, b, q, \bar q 
\end{eqnarray}
where $f^{\rm NLO}_i(x)$ is the parton density, evoluted 
at the NLO, relative to
the $i^{th}$ initial state particle, carrying a fraction $x$ of
the proton (or antiproton) longitudinal momentum
\begin{eqnarray}
f_i^{\rm NLO}(x) = f_i^{\rm (0)}(x)
 + {\alpha_s} f_i^{\rm (1)}(x)\,,
\end{eqnarray}
and ${\rm d} \hat \sigma^{\rm NLO}_{ij}$ are the differential cross
sections corresponding to the subprocesses given in Figs.
\ref{fig:nlovirt} and  \ref{fig:nloreal} computed at the 
one loop accuracy in QCD
\begin{eqnarray}
{\rm d} \hat \sigma^{\rm NLO}_{ij} = 
{\rm d} \hat \sigma^{\rm (0)}_{ij}+ { \alpha_s}
{\rm d} \hat \sigma^{\rm (1)}_{ij}\,.
\end{eqnarray}
The separation between virtual and real  
contributions in {${\rm d} \hat \sigma^{\rm (1)}_{ij}$} is performed 
in MCFM with the help of a Dipole Formalism \cite{Catani96}.

It is worth discussing the relative size of all the 
terms appearing in Eq. \ref{master}.
Setting the scale of the hard scattering to $\mu= M_Z$, the distribution
function $f_b^{\rm (0)}(x)$ intrinsically sums up 
all contributions of the order $\alpha_s^k L^k$ \cite{maltoni:hb},
where $L = \ln(\mu/m_b)$ is the large collinear logarithm
associated to the fact that $f_b^{\rm (0)}(x)$ describes
an exactly collinear gluon splitting $g \to b \bar b$.
Therefore, the leading order term
\begin{eqnarray}
\label{leadorda}
{\rm d} \sigma^{(0)} = \int_0^1 dx_1 \int_0^1 dx_2~
     f^{\rm (0)}_g{(x_1)} f^{\rm (0)}_b{(x_2)}~ 
      {\rm d} \hat \sigma_{gb}^{\rm (0)}~~+~~(g \leftrightarrow b)\,,
\end{eqnarray}
coming from the diagrams in Fig. \ref{fig:02} contains all possible
$\alpha_s^k L^k$ contributions
\begin{eqnarray}
\label{leadordb}
{\rm d} \sigma^{(0)} = \sum_{k=1}^{\infty} c_k \,\alpha_s^k L^k\,.
\end{eqnarray}
Including the diagrams in Figs. \ref{fig:nlovirt},
\ref{fig:nloreal}(a) and \ref{fig:nloreal}(b) takes into account
all corrections order $\alpha_s^{(k+1)} L^k$, because the corresponding
subprocesses multiply $f^{\rm (0)}_b{(x)}$ in Eq. \ref{master}. The 
corrections given in Fig. \ref{fig:nloreal}(c) correspond to the
contributions ${\cal O}(\alpha_s)$ or ${\cal O}(\alpha_s L) $, depending 
whether the final state $\bar b$ is or is not
collinear to one of the initial state gluons
\footnote{Note that such collinear or almost collinear  
configurations contribute to the exclusive process 
we are studying when  the final state  $\bar b$ 
is lost in the forward or backward regions of the detector.}.
They correct the leading order picture 
implicit in Eq. \ref{leadorda} of exactly collinearly produced $b$:
now the initial state $b$ can acquire a $p_t$.
However, as discussed, {\em all} collinear contributions  
are already included in the LO process, in particular the  $k= 1$
term in Eq. \ref{leadordb}, therefore we are facing an apparent 
double counting.
The key for understanding that this is not the case is noticing 
that other contributions
are present in Eq. \ref{master}, that correspond to the evolution
of the parton densities
\begin{eqnarray} 
      \alpha_s\, \left(f^{\rm (1)}_g{(x_1)} f^{\rm (0)}_b{(x_2)} 
              + f^{\rm (0)}_g{(x_1)} f^{\rm (1)}_b{(x_2)}\right)\, 
        {\rm d} \hat \sigma_{gb}^{\rm (0)}
~~+~~(g \leftrightarrow b)\,.
\end{eqnarray} 
Their effect is, among others, subtracting the 
$c_1 \alpha_s L$ term from the LO contribution
in Eq. \ref{leadordb}, so that
adding the corrections of Fig. \ref{fig:nloreal}(c), does not
imply double counting. The structure described so far 
is implemented in MCFM, that
we had to modify in order to disentangle 
$b$ and $\bar b$ production, which is necessary, in our case, 
for computing $A^b_{\rm FB}$.

Finally, we computed the pure  ${\cal O}(\alpha_s)$ corrections in Fig.
\ref{fig:nloreal}(d) with the help of ALPGEN.
In the next section we discuss our numerical findings. 
\section{Numerical results}
\begin{figure}[ht]
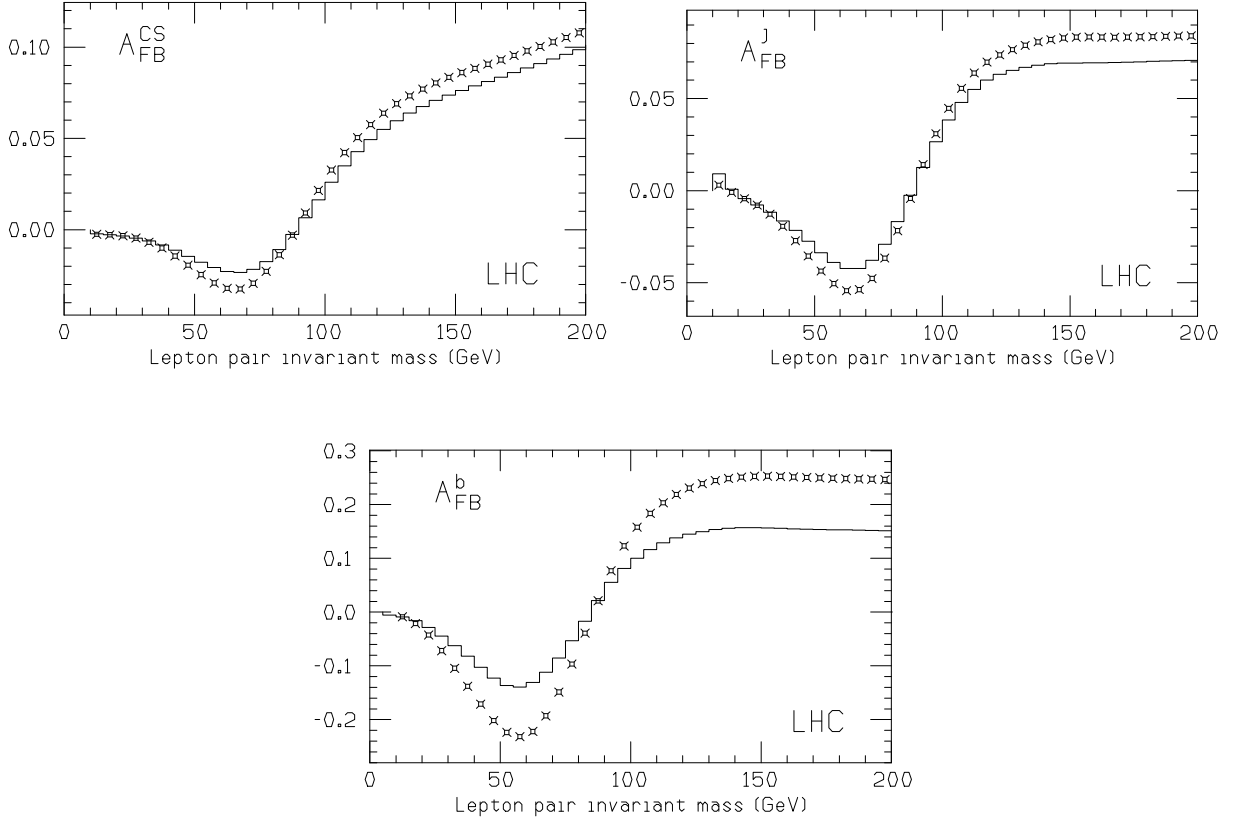

\begin{center}
\epsfig{file=Acs_lhc_conl.eps,height=8.cm,angle=90}
\epsfig{file=Aj_lhc_conl.eps,height=8.cm,angle=90}

\vskip 0.9cm

\epsfig{file=Ab_lhc_conl.eps,height=8.cm,angle=90}
\end{center}
\caption{\label{fig:1} NLO (solid histogram)
and LO (points) asymmetries at LHC.}
\end{figure}

\begin{figure}[ht]
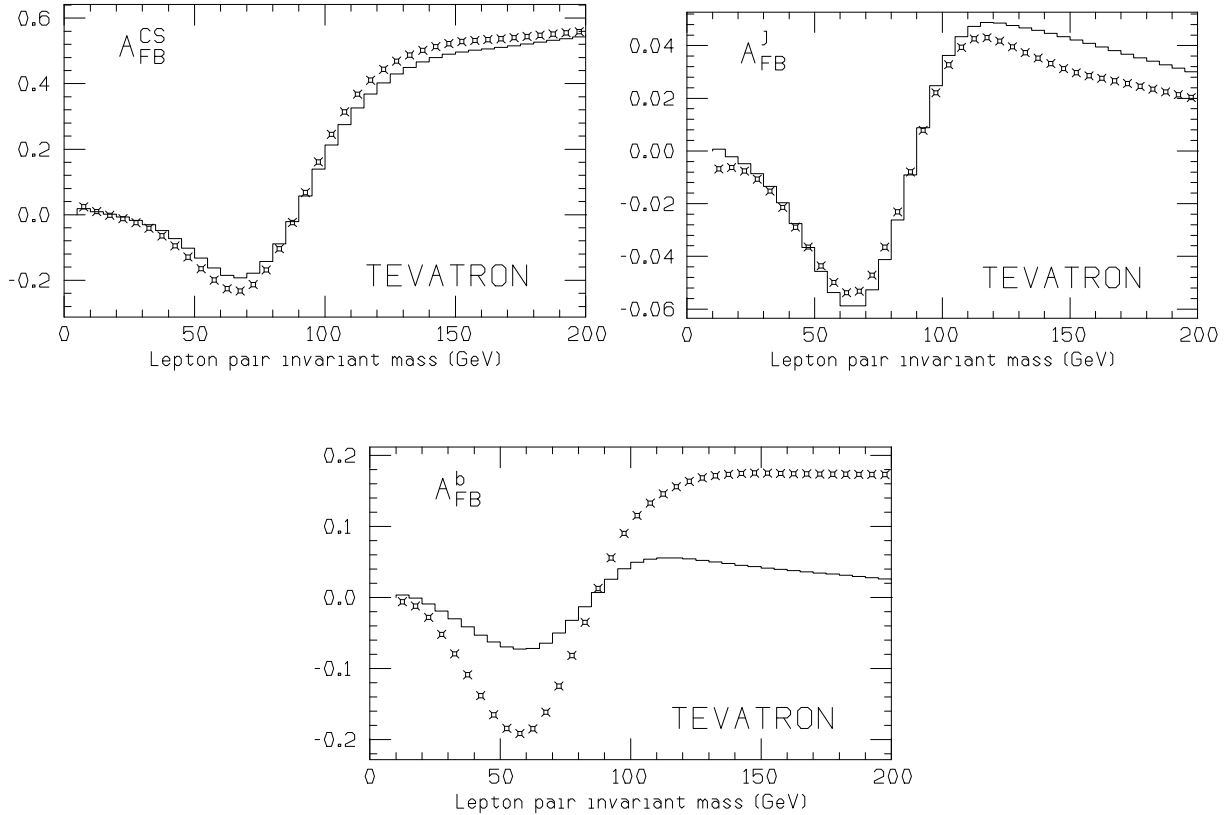

\begin{center}
\epsfig{file=Acs_tev_conl.eps,height=8cm,angle=90}
\epsfig{file=Aj_tev_conl.eps,height=8cm,angle=90}

\vskip 0.9cm

\epsfig{file=Ab_tev_conl.eps,height=8cm,angle=90}
\end{center}
\caption{\label{fig:2} NLO (solid histogram) and LO (points) 
asymmetries at Tevatron.}
\end{figure}

We present our numerical results for 
$e^-e^+j$ and $e^-e^+b$ at LHC and Tevatron in turn.
Our simulation of the set up at LHC {(Tevatron)} is as follows
\begin{eqnarray}
& p^e_t >20 \,{\rm GeV}, \,\,\, &
p^j_t >50\, { (30)} \,{\rm GeV}, \,\,\, 
\nonumber \\
& |\eta ^{e,j}| <2.5, \,\,\, &
\, \, \, \Delta R_{e,j} >0.4\,. \nonumber
\end{eqnarray}
For muon pairs the main difference would be the pseudorapidity 
coverage \cite{DetectorsAtlas, DetectorsCMS}.
We use the {\tt cteq6l1} ({\tt cteq6m}) parton distributions
at LO (NLO) \cite{cteq}.
The effect of smearing the lepton and jet energies has been 
studied at LO and found to be negligible \cite{Aguila02}, 
therefore we do not include it here.
On the other hand, the dominant background processes are 
expected to be the same as for Drell-Yan production, 
namely, jets misidentified as $e^{\pm}$, and
$p \ppbar \to W^+ W^- j \to e^+ e^- \nu_e \bar \nu_e j$.
They are understood experimentally, at least at Tevatron, and can 
be considered under control \cite {CDF05a}.
The LO and NLO production rates at LHC and Tevatron are given in 
Table \ref{table:2}.
\begin{table*}[ht]
\begin{center}
\begin{tabular}{|c||c|c|c|c|}
\hline
{\rm Contributing} &  \multicolumn{2}{c|}{\rm LHC}
  & \multicolumn{2}{c|}{\rm Tevatron} \\
{\rm process} & \multicolumn{1}{c}{\rm LO} & \multicolumn{1}{c|}{\rm NLO}
  & \multicolumn{1}{c}{\rm LO} & \multicolumn{1}{c|}{\rm NLO}\\
\hline
\hline
$g \qqbar \rightarrow Vj(j)$ & 44.3 & 53.4 
  & 3.40 & 4.77 \\
$\begin{array}{c}
q\bar q \rightarrow Vj(j) \\
\qqbar \qqbar \rightarrow Vj(j) \\
g g \rightarrow Vj(j) \\
\end{array}$ & 
$\begin{array}{c}
8.4 \\
- \\
- \\
\end{array}$ & 
$\left.
\begin{array}{c}
 \\
 \\
 \\
\end{array}
\right\} 3.7$ 
& $\begin{array}{c}
  4.61 \\
  - \\
  - \\
  \end{array}$ & 
  $\left.
  \begin{array}{c}
   \\
   \\
   \\
  \end{array}
  \right\} 2.76$ \\
\hline
{\rm Total} & 52.7 & 57.1 
  & 8.01 & 7.53 \\
\hline
\hline
$\begin{array}{c}
gb \rightarrow Vb(g) \\
gg \rightarrow Vb(\bar b) \\
\qqbar b \rightarrow Vb(\,\,\qqbar) \\
\end{array}$ &
$\begin{array}{c}
1.81 \\
- \\
- \\
\end{array}$ &
$\left.
\begin{array}{c}
 \\
 \\
 \\
\end{array}
\right\} 1.81 $ 
  & $\begin{array}{c}
  0.038 \\
  - \\
  - \\
  \end{array}$ &
  $\left.
  \begin{array}{c}
   \\
   \\
   \\
  \end{array}
  \right\} 0.049 $ \\
$q\bar q \rightarrow Vb(\bar b)$ & $-$ & 0.06 
  & $-$ & 0.025 \\
\hline
{\rm Total} & 1.81 & 1.87 
  & 0.038 & 0.074 \\
\hline
\end{tabular}
\end{center}
\caption{\label{table:2}
Estimates for the $e^-e^+j$ and $e^-e^+ b$ 
cross sections at LHC ($\sqrt s = 14$ TeV)
and Tevatron ($\sqrt s = 1.96$ TeV) in pb. 
The jet transverse momenta are required to be larger than 
50 (30) GeV at LHC (Tevatron)
and all pseudorapidities $|\eta |$ smaller than 2.5. 
The $p_t$ of the leptons is larger than 20 GeV.
The separations in the pseudorapidity-azimuthal angle 
plane satisfy $\Delta R > 0.4$ 
and $M_{e^-e^+}$ is within the range $[75,105]\, {\rm GeV}$.
$\qqbar$ means summing over $q$ and $\bar q$  contributions.}
\end{table*}     
In Figs. \ref{fig:1} (\ref{fig:2}) we show
the corresponding charge asymmetries, ${\rm A}^{\rm CS}_{\rm FB}$ 
relative to the initial parton 
and ${\rm A}^j_{\rm FB}$, ${\rm A}^b_{\rm FB}$ relative to the final jet. 

The effect of the ${\cal O}(\alpha_s)$ corrections is moderate for
${\rm A}^{\rm CS}_{\rm FB}$ and ${\rm A}^j_{\rm FB}$, 
but sizable for ${\rm A}^b_{\rm FB}$, especially
at Tevatron. This is basically due to the genuine new higher order process
$q \bar q \to  V b (\bar b)$
in Fig. \ref{fig:nloreal}(d), that tends to wash out the asymmetry 
which is mainly associated to the (a) contribution in Fig. \ref{fig:02}. 
This change is more pronounced at Tevatron energies, where the
$q \bar q $ content of the (anti)proton is larger
(see Table \ref{table:2}). 
In all cases the asymmetries at NLO diminish, except for 
${\rm A}^j_{\rm FB}$ at Tevatron where this asymmetry is smaller.

Near the $Z$ pole, $M_{e^-e^+}\sim M_Z$,  
the asymmetries can be approximated 
by \cite{Rosner89}
\be
{\rm A} =
{\rm b}({\rm a}-\sin ^2 \theta ^{\rm lept}_{\rm eff} (M_Z^2)),
\label{appr.asymm}
\en 
translating then their measurement into a precise 
determination of 
$\sin ^2 \theta ^{\rm lept}_{\rm eff} (M_Z^2)$. 
In Table \ref{table:3} we collect the asymmetry estimates, their 
statistical precision, the cross sections and the 
precision reach 
$\delta \sin ^2 \theta _{\rm eff}^{\rm lept}$ of LHC 
and Tevatron for $M_{e^-e^+}$ in the range $[75,105]\, {\rm GeV}$, 
by assuming an integrated Luminosity  {$L$} of
100 {(10)} $fb^{-1}$ at LHC {(Tevatron)}.
\begin{table*}[ht]
\begin{center}
\begin{tabular}{|c|r|c|c|c|c|c|}
\hline
\begin{tabular}{c}
LO \vspace{-0.3cm} 
 \\ NLO
\end{tabular}
& $\sigma({\rm pb})~~~~$ &   &  ${\rm A}_{\rm FB}$ & 
$\delta{\rm A}_{\rm FB}$ & $\delta \sin ^2 \theta _{\rm eff}^{\rm lept}$ \\
\hline
\hline
LHC      & $\sigma^{Vj}=$ 53  & $A^{\rm CS}_{\rm FB}$ & $8.7 \times 10^{-3}$ & $4.4 \times 10^{-4}$ & $1.3 \times 10^{-3}$ \\ 
         &      {  57} &          & ${  6.8 \times 10^{-3}}$ & ${  4.2 \times 10^{-4}}$ & ${  1.3 \times 10^{-3}}$ \\
         &                    & $A^{j}_{\rm FB}$  & $1.2 \times 10^{-2}$ & $4.4 \times 10^{-4}$ & $8.8 \times 10^{-4}$ \\
         &               &          & ${  1.1 \times 10^{-2}}$ & ${  4.2 \times 10^{-4}}$ & ${  1.1 \times 10^{-3}}$ \\
         & $\sigma^{Vb}=$ 1.8 & $A^{b}_{\rm FB}$  & $7.5 \times 10^{-2}$ & $2.3 \times 10^{-3}$ & $8.7 \times 10^{-4}$ \\ 
         &          {  1.9}&          & ${  4.9 \times 10^{-2}}$ & ${  2.3 \times 10^{-3}}$ & ${  1.4 \times 10^{-3}}$ \\ 
       \hline \hline
Tevatron & $\sigma^{Vj}=$ 8.0 & $A^{\rm CS}_{\rm FB}$ & $6.4 \times 10^{-2}$ & $3.5 \times 10^{-3}$        & $1.4 \times 10^{-3}$ \\ 
         &          {  7.5}&   & ${  5.5 \times 10^{-2}}$ & ${  3.6 \times 10^{-3}}$ & ${  1.7 \times 10^{-3}}$ \\
         &                    & $A^{j}_{\rm FB}$  & $      9.9 \times 10^{-3}$ & $3.5 \times 10^{-3}$ &        $8.1 \times 10^{-3}$ \\
         &                    &   & ${  1.1 \times 10^{-2}}$ & ${  3.6 \times 10^{-3}}$ & ${  7.2 \times 10^{-3}}$ \\
         & $\sigma^{Vb}=$ 0.04& $A^{b}_{\rm FB}$  & $      5.5 \times 10^{-2}$ & $5.1 \times 10^{-2}$ &        $2.5 \times 10^{-2}$ \\ 
         &         {  0.07}&   & ${  2.7 \times 10^{-2}}$ & ${  3.7 \times 10^{-2}}$ & ${  4.7 \times 10^{-2}}$ \\
\hline
\end{tabular}
\end{center}
\caption{\label{table:3} Estimates 
for the $e^-e^+j$ and $e^-e^+ b$ 
cross sections and asymmetries defined in the text
with $M_{e^-e^+}$ in the range $[75,105]\, {\rm GeV}$.
The first row of each entry is the LO result, while the second one 
refers to the NLO. 
The integrated luminosity as well as the cuts can be found
in the text.
The statistical precisions are also given, to be compared with 
the current effective weak mixing angle uncertainties  
at LEP and SLD from asymmetries only 
$1.6\times 10^{-4}$, and from ${\rm A}_{\rm FB}^b$ at the 
$Z$ pole $2.9\times 10^{-4}$
\cite{LEP02}.}
\end{table*}     
In the Table we assumed a $b$-tagging 
efficiency $\epsilon$ of $100\ \%$, no contamination 
$\omega$ and, in particular, no charge misidentification. 
The statistical precisions $\delta {\rm A}$ and
$\delta \sin ^2 \theta _{\rm eff}^{\rm lept}$ 
are proportional to $\epsilon ^{-\frac{1}{2}}$, 
and the asymmetries A to $1-2\omega$. 
Therefore the contamination multiplies 
$\delta \sin ^2 \theta _{\rm eff}^{\rm lept}$ 
by $(1-2\omega)^{-1}$.
Thus, if we only consider semileptonic $b$ decays, 
implying $\epsilon \sim 0.1$ and $\omega \sim 0$, 
$\delta {\rm A}$ and
$\delta \sin ^2 \theta _{\rm eff}^{\rm lept}$ increase 
by a factor $\sim 3$.  
In practice we must try to maximize the quality factor 
${\rm Q} = \epsilon (1-2\omega)^2$ \cite{BaBar}. 
Although the exact value of $\omega$ can only be inferred
from dedicated experimental studies, it is instructive to have
an idea of its typical size. This can be achieved by
comparing the charge separation $\delta_b$, measured at LEP 
using a jet charge technique, with the value $2 Q_b = -\frac{2}{3}$
one would get if the $b$ quark could be directly observed
\cite{Abreu:1998nr}.
This comparison gives $\omega \sim 0.3$.
The statistical precisions given in Table \ref{table:3} are 
certainly optimistic for systematic errors 
are also sizable. At any rate  approaching the quoted precisions 
will be an experimental challenge.

A second  source of uncertainty, that is not accounted for in
Table \ref{table:3}, is the dependence of the results
on the chosen set of parton densities. 
We investigated it by recomputing asymmetries and statistical precisions 
using different parton distribution sets in the classes {\tt cteq} and {\tt mrst} 
\cite{mrst}.
By doing so, variations of the asymmetries of the order of 10\% 
can be easily 
observed  in the range $75~{\rm GeV} < M_{e^-e^+} < 105~{\rm GeV}$, 
both at Tevatron and LHC, while the statistical precisions 
are not significantly affected.
This rather important dependence on the parton densities can be 
considered as an extra handle provided by the asymmetry measurements 
in constraining the parton distribution functions.
Conversely, with a more precise knowledge of them, the charge 
asymmetries can be used for precision measurements.
\section{Conclusions}
In summary, the large $Vj$ production 
cross section at hadron colliders and the possibility of 
measuring the lepton asymmetries relative to the final jet 
allow for a precise determination of the effective electroweak 
mixing angle. We have evaluated them to NLO, confirming to a large
extent the LO results. If there is an efficient $b$-tagging and charge 
identification, these events with a $b$ jet also allow for 
a new determination of ${\rm A}_{\rm FB}^b$. 
The corresponding statistical precisions are collected
in Table \ref{table:3}. As in Drell-Yan production \cite{Rosner96}, 
these processes can be also sensitive to new physics for 
large $M_{e^-e^+}$, especially to new gauge bosons.  
\section{Acknowledgments}
We thank A. Bueno, J.M. Campbell and T. Rodrigo for useful comments.
This work was supported in part by MCYT under 
contract FPA2003-09298-C02-01 and by 
Junta de Andaluc{\'\i}a group FQM 101, and by
MIUR under contract 2004021808\_009.

\vfill

\end{document}